# Bluetooth Data Exchange Between Android Phones Without Pairing

Yakov Shafranovich[1]

Noom, Inc.
yakov@noom.com

**Abstract:** In this paper we describe a novel method of exchanging data between Bluetooth smartphones on the Android platform without requiring pairing between devices. We discuss our approach of encoding and decoding data inside the UUIDs used by the Bluetooth Service Discovery Protocol (SDP). Future research remains to be done on the latency, bandwidth and compatibility of this approach, as well as the possibility of utilizing other protocols in conjunction with this method.

## 1. Introduction.

The problem that we are trying to solve sound simple on the surface – we want to figure out how to get information exchanged between multiple Android smartphones in close proximity to each other, but with the least amount of barriers to the users of those devices. Think of a scenario like a concert, a public protest or a party where many people are attending, and want to exchange data with each other. Add the possibility of unreliable wireless access or saturated mobile networks, or perhaps networks that are actively blocking activity of their users. A peer to peer solution for data exchange would be a better alternative in such cases and our approach, as discussed in this paper, offers one such solution.

### 1.1. Existing Work.

There are several protocols available on smartphones that can potentially be used to facilitate peer to peer ad-hoc communication. These include protocols like XMPP [1], WiFi [2] and WiFi Direct [3], Bluetooth classic [4] and LE [5], as well as more novel approaches like sound [6]. Many have potential problems including privacy issues [7] such as the necessity to reveal phone numbers and MAC addresses, security issues once devices are paired, possibilities of monitoring or blocking, etc.

There also exist several novel ways of using Bluetooth to solve this problem, similar to the approach described in this paper. One possibility is to use the device name string to exchange information [8], but this would be visible to end users. Another approach [9] is to use the UUID strings to exchange user identifiers and then use a secondary channel to exchange messages.

[1] I want to thank all of my colleagues at Noom, Inc. for helping with this project.

# 2. Encoding Data in Bluetooth UUIDs.

For our solution, we choose to use Bluetooth since it allows devices located in proximity to each other to communicate, and is supported by most Android smartphones. We also choose to use a novel method to exchange data via service identifiers, in order to avoid the requirement for the devices to pair, in order to reduce the friction among users and to increase security.

## 2.1. Overview of Bluetooth Protocol.

Bluetooth is a wireless protocol that allows communications between wireless devices located within close proximity. Bluetooth devices are uniquely identified using MAC IDs, following a similar scheme as network card addresses [10]. Depending on the class of the device, the range can be anywhere from 10 to 100 meters [11].

Every device also offers a device name, and a list of supported services. The device name is a simple string, limited to 248 bytes [10]. Services are identified using UUID numbers. In the classical Bluetooth protocol scheme, devices use the Bluetooth Service Discovery Protocol (SDP) to scan for nearby devices and discover what services they offer. While applications can use their own UUIDs to identify services, the Bluetooth SIG maintains a list of standard “well known” services.

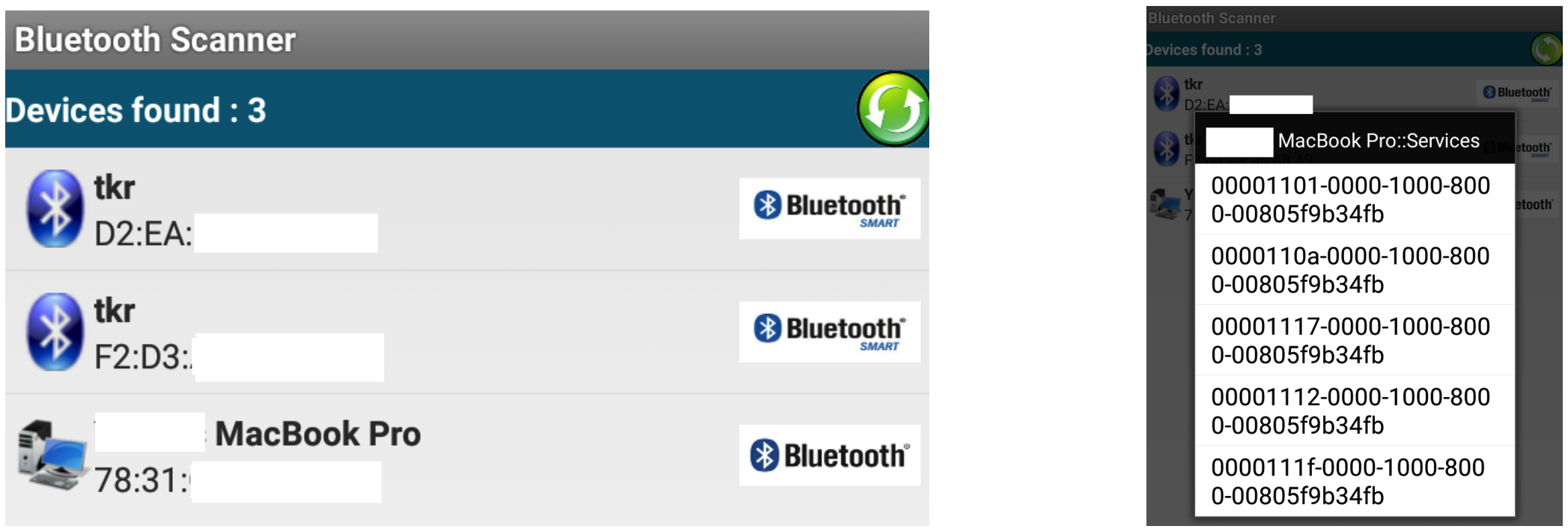

**Figure 1. Example of Bluetooth devices and services on an Android phone.**

Besides classic Bluetooth, a recent development that appears in Bluetooth protocol v4.0, is Bluetooth Low Energy, which allows devices such as sensors, beacons, and others to exchange data while consuming little power [10]. While Bluetooth LE can also be used to communicate in a similar fashion, we choose not to use BLE because it is not supported well on older devices.

## 2.2. Overview of UUIDs.

UUIDs are standardized in [12]. They are 128-bit unique identifiers that can be generated autonomously, which makes them useful for replication and other areas of computing. Our interest in UUIDs is related to the fact that they can be used to carry arbitrary binary data.

UUIDs are defined as containing 16 bytes or 128-bits. There are also 5 different version of UUIDs some of which constraint what type of data can appear. We are using version 4, which accommodates randomly generated data and provides the least amount of restrictions. UUIDs are usually formatted as 32 alphanumeric characters separated by hyphens as “8-4-4-4-12” characters or “4-2-2-2-6” in bytes.

# 3. Implementation Details.

In our sample implementation, we developed a chat program for Android smartphones that allows people located nearby to chat in a public chatroom. The messages they are putting into the chatroom are exposed on each device via Bluetooth service identifiers, and their device name is used as their chatroom handle. We continuously scan nearby devices to identify their latest messages and display those to the user. We also allow the user to change their own message.

Our approach solves the original problem of peer to peer ad-hoc communication with less barriers and more security by using a standard protocol build into smartphones, bypassing the need to pair the phones, and exchanging data between unpaired phones by avoiding the use of Bluetooth sockets between phones.

For our sample implementation, we a Nexus 4 phone running Android v5.1. All development was done on Android Studio v1.2.2 running on Mac OS v10.10.4.

## 3.1. How We Are Passing Data.

Every Bluetooth device advertises a list of services using UUIDs numbers. We will be storing information as fake Bluetooth services, by encoding binary data in the UUID strings representing available services. There are three advantages to this method:

1. Looking at available services for a device does not require pairing.
2. Because a single device can list many services, this increases the bandwidth available for data exchange.
3. Because this data is not readily visible to regular users and operating systems via GUI, it affords a small measure of privacy.

There are two constraints on version 4 UUIDs: the first requires setting one specific character to “4” to indicate the version being used, and second requiring another character to be set to “8” since it is reserved (for simplicity we are using the entire 4 bits). The rest of the UUID leaves us with 120 bits or 15 bytes to pass information around. The Bluetooth standards [10] further reserve any UUIDs using the values “xxxxxxxx-0000-1000-8000-00805F9B34FB”. This is not an issue for us since we are using version 4 UUIDs, unlike the reserved ones in Bluetooth that use version 1.

To distinguish our services from others, we will use the last 4 characters / 2 bytes as “C0DE”. This will leave us with 104 bits or 13 bytes per service UUID to use with data. It would be trivial to extend this to support multiple UUIDs with different messages in them.

```
xxxxxxxx-xxxx-4xxx-8xxx-xxxxxxxxC0DE
              |    |             |
              |  reserved      special flag
            version     for detecting our services
```

**Figure 2. Sample UUID string.**

## 3.2. Decoding and Encoding Data.

Detection of our UUIDs with Java is pretty simple, we can use a regular expression to match against the string form of the UUID.

```
String uuid = "ff724081-fe5d-4fb2-8745-af149cc2c0de";
String regex = “([a-f0-9]{8})-([a-f0-9]{4})-4([a-f0-9]{3})-8([a-f0-9]{3})-
([a-f0-9]{8})c0de”;
Pattern pattern = Pattern.compile(regex);
Matcher m = pattern.matcher(uuid.toLowerCase());
if (m.find()) {
    String hexData = m.group(1) + m.group(2) + m.group(3) + m.group(4) +
m.group(5);
    String binaryData = new BigInteger(hexData, 16).toString(2);
} else {
...
}
```

**Figure 3. Code for decoding data in UUIDs.**

To encode data, we would pad it with zeros, and convert to binary, then encode inside a UUID.

```
String data = "12345678901234";
String paddedData = String.format("%-14s", data).replace(' ', '\0');
String hexData =
     String.format("%028x", new BigInteger(1,
paddedData.getBytes()));
String uuidString =
hexData.substring(0, 8) + "-" +
hexData.substring(8, 12) + "-4" +
hexData.substring(12, 15) + "-8" +
hexData.substring(15, 18) + "-" +
hexData.substring(18) + "c0de";
```

**Figure 4. Code for encoding data in UUIDs.**

## 3.3. Registering Bluetooth services.

Once we have the data we want to encode, it is trivial to actually register the service as follows.

```
BluetoothAdapter mBluetoothAdapter =
BluetoothAdapter.getDefaultAdapter();
BluetoothServerSocket btSocket = mBluetoothAdapter.
       listenUsingInsecureRfcommWithServiceRecord("example", uuid)
```

**Figure 5. Code for registering a Bluetooth service.**

Because we only want the UUIDs, it does not matter whether secure or insecure RFCOMM is used. When changing the message, we would close the old socket and open a new one.

## 3.4. Finding Bluetooth Services

To locate our data, three steps would be needed. First, we would need to scan the surrounding area for Bluetooth devices and watch out for new devices as per [13]. Once a device has been found, we would call **fetchUuidsWithSdp()** [13] and then **getUuids()** [13] to get a list of UUIDs for that device. Third, we would go through the list of UUIDs and look for the ones marked with the special ending as described above, and extract the data.

```
private final BroadcastReceiver mReceiver = new
BroadcastReceiver() {
    public void onReceive(Context context, Intent intent) {
        String action = intent.getAction();
        if (BluetoothDevice.ACTION_FOUND.equals(action)) {
           BluetoothDevice device =

intent.getParcelableExtra(BluetoothDevice.EXTRA_DEVICE);
           device.fetchUuidsWithSdp();
        } else if (BluetoothDevice.ACTION_UUID.equals(action)) {
           BluetoothDevice device =

intent.getParcelableExtra(BluetoothDevice.EXTRA_DEVICE);
           Parcelable[] uuids =

intent.getParcelableArrayExtra(BluetoothDevice.EXTRA_UUID);
           for (Parcelable ep : uuids) {
              String uuid = ep.toString();
              // Detect/extract data here
           }
        }
    }
};

IntentFilter filter1 = new
IntentFilter(BluetoothDevice.ACTION_FOUND);
registerReceiver(mReceiver, filter1);

IntentFilter filter2 = new
IntentFilter(BluetoothDevice.ACTION_UUID);
registerReceiver(mReceiver, filter2);
```

**Figure 6. Code for finding Bluetooth services.**

# 4. Latency, Bandwidth and Security.

It remains unknown how often it would be possible to change the UUIDs for the device and have that information relayed in a timely fashion. In our testing, we were able to change the message with less than an one minute interval, but we did not have the opportunity to test this approach on a wide range of hardware devices. Additionally,  the scanning of Bluetooth devices and enumeration of their services is a longer operation than simply looking at services for an already known or paired device. This may be another area to explore for future optimization.

Bandwidth in our implementation was 13 bytes per each UUID. Based on our testing, we were limited to opening 7 additional sockets on an Android device, at a single time for a total of 13×7 = 91 bytes. For scanning, we were limited to 21 service records per device, for a total of 273 bytes, which isn't much more than the approach described by [8]. For practical implementations, a hybrid approach as outlined in [9] maybe more appropriate where UUID encoding is used to share user identifiers and the actual message content is exchanged using another channel such. However, because these restrictions may be due to the specific hardware and software configuration we tested, it remains to be explored whether different Android devices may in fact have more bandwidth for this method.

In regards to security, exchanging data via unpaired Bluetooth devices provides a greater measure of security since the devices are only exposing data via service identifiers and device names. In contrast, paired devices can open arbitrary sockets and exchange files, which exposes each of the devices to a greater possibility of hacking. We also choose to forgo encryption in order to simplify our implementation, but the use of encryption would certainly improve the security of our approach.

# 5. Future Areas for Research

Some possible areas of research in the future would be looking at Bluetooth and other protocols to see if they can be used to facilitate similar types of communications. WiFi Direct [3] and Bluetooth LE [5] may also be a possible avenue to explore.

From a security perspective, the approach described in this paper if paired with vulnerability in the Bluetooth stack, may be used to propagate Bluetooth viruses between smartphones, which can form a self updating mesh network like the designed outlined in [14]. The use of encryption also remains to be explored in conjunction with this method. These and other security issues are good areas for future research.

# References

1. Saint-Andre, P., Ed., "Extensible Messaging and Presence Protocol (XMPP): Core", RFC 3920, DOI 10.17487/RFC3920, October 2004, <http://www.rfc-editor.org/info/rfc3920>
2. "The Serval Project." Accessed July 2, 2015. <http://developer.servalproject.org/>
3. D. Camps-Mur, A. Garcia-Saavedra, and P. Serrano, "Device to device communications with wifi direct: overview and experimentation," IEEE Wireless Communications Magazine, 2012
4. "Open Garden Apps." Accessed July 2, 2015. <https://opengarden.com/apps>
5. "CSRMesh." Accessed July 2, 2015. <https://wiki.csr.com/wiki/CSRmesh>
6. "Chirp Technology: An Introduction." Chirp.io. Accessed July 2, 2015. <http://chirp.io/technology/>
7. Citizen Lab, "Asia Chats: Update on Line, KakaoTalk, and FireChat in China," Citizen Lab Research Brief No. 44, July 2014. <https://citizenlab.org/2014/07/asia-chats-update-line-kakaotalk-firechat-china/>
8. Cohen, Joseph P.. "Wireless Message Dissemination via Selective Relay over Bluetooth (MDSRoB)." July 30, 2013. <http://arxiv.org/abs/1307.7814>

9. Newbry, Joe S., "Samantha: A Social Location-Based Framework for iOS Applications" (2014). CMC Senior Theses. Paper 939. <http://scholarship.claremont.edu/cmc_theses/939>
10. “Bluetooth Specification version 4.2.” Bluetooth SIG. December 2, 2014. <https://www.bluetooth.org/en-us/specification/adopted-specifications>
11. Wright, Joshua. "Dispelling Common Bluetooth Misconceptions." SANS Institute. September 19, 2007. <https://www.sans.edu/research/security-laboratory/article/bluetooth>
12. Leach, P., Mealling, M., and R. Salz, "A Universally Unique IDentifier (UUID) URN Namespace", RFC 4122, DOI 10.17487/RFC4122, July 2005, <http://www.rfc-editor.org/info/rfc4122>.
13. “Android Developer Guide: Bluetooth”. Google, Inc. Accessed July 2, 2015. <https://developer.android.com/guide/topics/connectivity/bluetooth.html>
14. Wily, Brandon., “Curious Yellow: The First Coordinated Worm Design.” Accessed July 2, 2015. <http://blanu.net/curious_yellow.html>